\documentclass[%
 reprint,
 amsmath,amssymb,
 aps,
]{revtex4-1}

\usepackage{subcaption}
\usepackage{graphicx}
\usepackage{dcolumn}
\usepackage{bm}

\usepackage{xr}
\begin{document}


\title{Anomalous Hall Effect in type-I Weyl metals}

\author{J. F. Steiner$^{1,2}$, A. V. Andreev$^2$, D. A. Pesin$^3$}
\affiliation{%
 $^1$ITP, Heidelberg University, Philosophenweg 12, 69120 Heidelberg, Germany\\
 $^2$Department of Physics, University of Washington, Seattle,  WA 98195, USA\\
$^3$Department of Physics and Astronomy, University of Utah, Salt Lake City, UT 84112, USA
}%

\date{\today}

\begin{abstract}
We study the \emph{ac} anomalous Hall conductivity $\sigma_{xy}(\omega)$ of a Weyl semimetal with broken time-reversal symmetry. Even in the absence of free carriers these materials exhibit a ``universal'' anomalous Hall response determined solely by the locations of the Weyl nodes.
We show that the free carriers, which are generically present in an undoped Weyl semimetal, give an additional contribution to the ac Hall conductivity. We elucidate the physical mechanism of the effect and develop a microscopic theory of the free carrier contribution to $\sigma_{xy}(\omega)$. The latter can be expressed in terms of a small number of parameters (the electron velocity matrix, the Fermi energy $\mu$, and the ``tilt" of the Weyl cone). The resulting $\sigma_{xy}(\omega)$ has resonant features at $\omega \sim 2 \mu$ which may be used to separate the free carrier response from the filled-band response using, for example, Kerr effect measurements. This may serve as diagnostic tool to characterize the doping of individual valleys.
\end{abstract}

\pacs{Valid PacS appear here}%

\maketitle


Weyl semimetals (WSMs) are topologically-nontrivial conductors in which the spin-nondegenerate valence and conduction bands touch at isolated points in the Brillouin zone, the so called ``Weyl nodes''~\cite{Herring,WanVishwanath2011,BurkovBalents2011,Ran2011,Kane2012,Wang2012,Wang2013,BernevigDai2015,Turner2013,Hosur2013}. The electron spectrum near the nodes is described by the chiral Weyl Hamiltonian, see Eq.~\eqref{Hamilton} below. The nodes occur in pairs of opposite chirality \cite{nielsen81}. The WSM phase requires either time reversal (TR) or inversion (I) symmetry (or both) to be broken \cite{murakami07}. Recently experimental evidence for the WSM phase was reported in non-centrosymmetric TaAs~\cite{discovery2TaAs15,Xu2015,Lv2015,Lv2015_2,Yang2015,Xu2015_2} and a range of other I-breaking materials \cite{wsms1,wsms2,wsms3,wsms4,wsms5,wsms6}.
TR-breaking WSMs have not been found yet but there are several promising candidates \cite{borisenko15,trbreakingCandidate1,trbreakingCandidate2,trbreakingCandidate3}.

In the situation where the touching valence/conduction bands are completely filled/empty,
such TR-breaking WSMs were shown to exhibit an anomalous Hall effect (AHE)~\cite{Haldane2004,Ran2011,aheMultilayer11,zyuzin12} that is ``universal'' in the sense that it only depends on the location of the nodes in the Brillouin zone (BZ). The theory of this contribution, below referred to as $\sigma_{xy}^{\textrm{(band)}}(\omega)$, was extended to the \textit{ac} regime \cite{kargarianKerr15}.

\begin{figure}[t!]
\centering
\includegraphics[width=0.95\columnwidth]{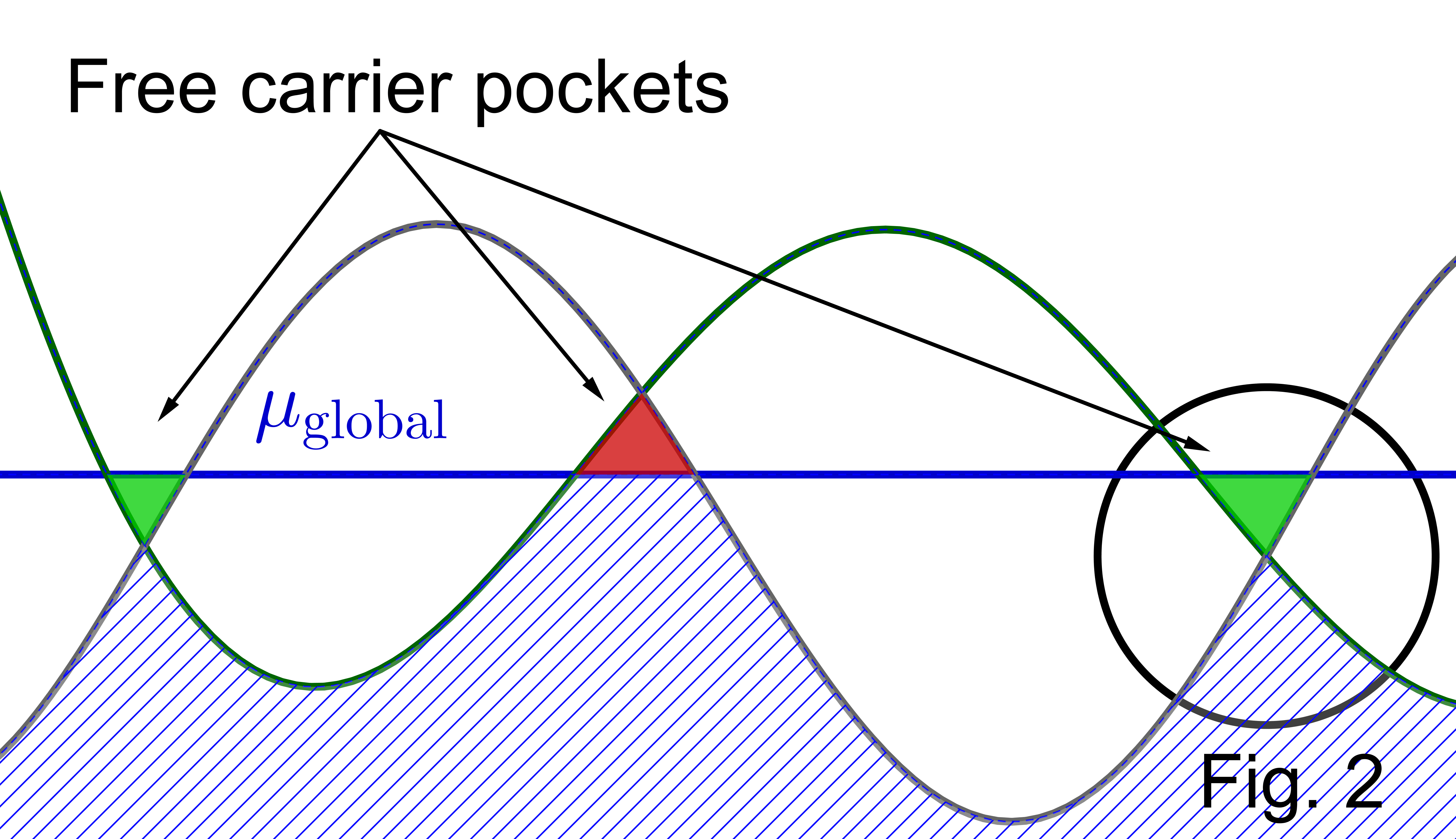}
\caption{\footnotesize Projection of generic band structure with Weyl points close to the chemical potential.
Free carriers are localized to electron (green) or hole (red) pockets near the Weyl nodes.
}
\label{bandsketch1}
\end{figure}

However, in a generic WSM, including all presently discovered ones,  free carriers of both electron and hole type are present, see Fig.~\ref{bandsketch1}.
In this work we develop a microscopic theory of the \textit{ac} anomalous Hall effect in a generic WSM. We show that the free carriers present near the nodal points provide a distinct contribution to the AH conductivity $\sigma^{\textrm{(free)}}_{xy} (\omega)$.

This \textit{free carrier} contribution has a resonant structure at frequencies on the scale of the Fermi energy of the free carriers. This feature should manifest itself in the spectrum of the magneto-optical Kerr effect and may find application as a diagnostic tool for materials characterization of WSMs.

In order to separate the free carrier contribution from that of the the filled bands  (see Fig.~\ref{bandsketch1}),
\begin{equation}\label{eq:sigma_sum}
  \sigma_{\alpha\beta}(\omega)=\sigma^{\textrm{(band)}}_{\alpha\beta}(\omega) + \sigma^{\textrm{(free)}}_{\alpha\beta}(\omega),
\end{equation}
we note that in optical response the external electric field couples only electron states with equal quasimomentum.
As a result the Kubo formula for the optical conductivity may be expressed as a sum over quasimomenta
\begin{equation}\label{eq:sigma_free}
  \sigma^{\textrm{(free)}}_{\alpha\beta}(\omega)= \sum_{\bm{p}}\sigma_{\alpha\beta}(\omega,\bm{p})\left[n_{+}(\bm{p})-n_{-}(\bm{p})+1\right],
\end{equation}
where $n_{\pm}(\bm{p})$ denotes the Fermi occupation function in the conduction/valence band, and by $\sigma_{\alpha\beta}(\omega,\bm{p})$ we denote the matrix elements and energy denominators. The filled band contribution in Eq.~\eqref{eq:sigma_sum} has the form, $\sigma^{\textrm{(band)}}_{\alpha\beta}(\omega)=\sum_{\bm{p}}\sigma_{\alpha\beta}(\omega,\bm{p})\times(-1)$.

\begin{figure}[t!]
\centering
\includegraphics[width=0.95\columnwidth]{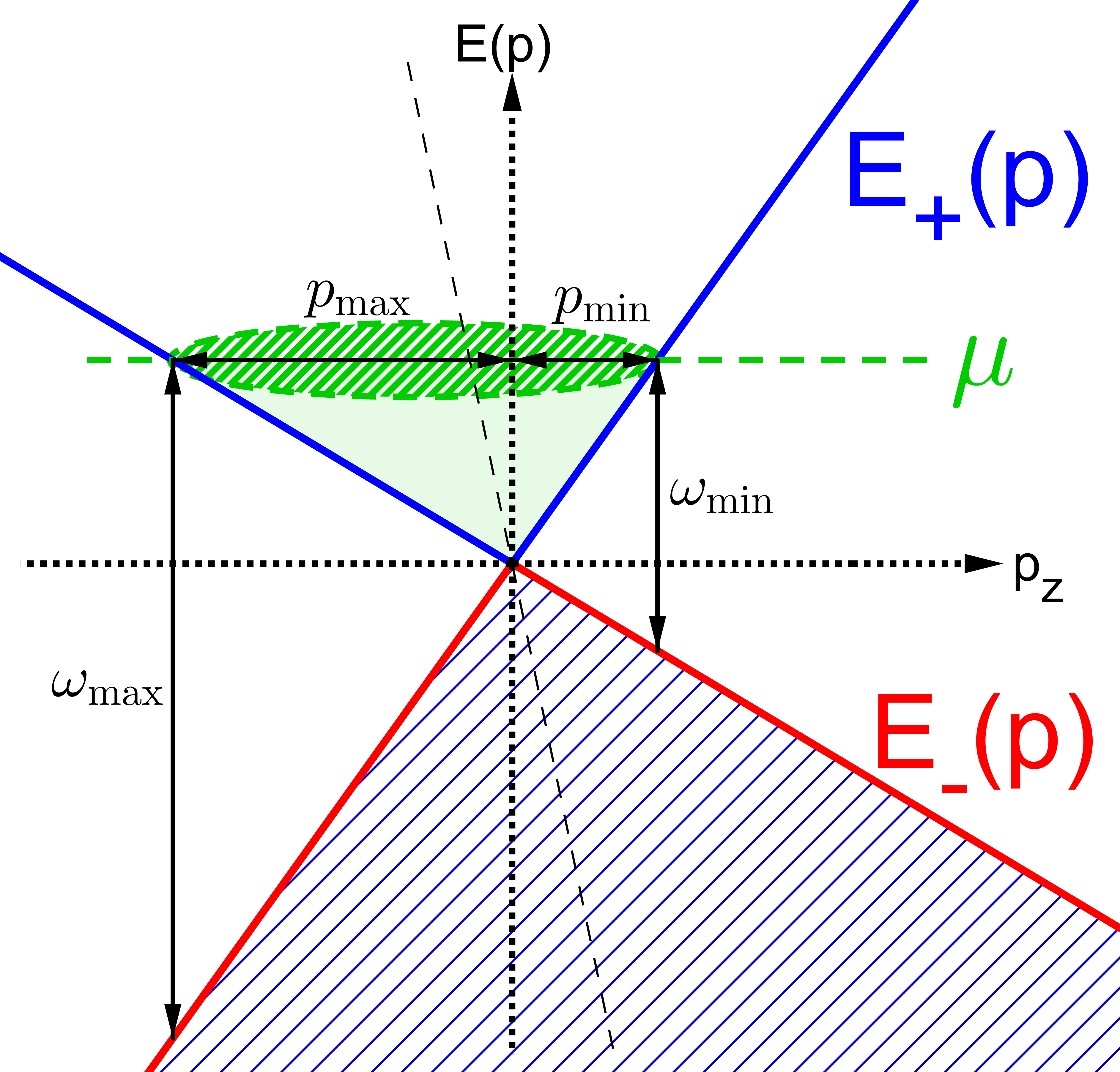}
\caption{\footnotesize Projection of the dispersion close to a generic Weyl node for $\bm{u}\vert\vert\hat{p}_z$ and $v_{ij}=v_f\delta_{ij}$. The Fermi-surface forms an ellipse centered away from the nodal point. The occupation factor $\left[n_+(\bm{p})-n_-(\bm{p})+1\right]$ is non-zero whenever transitions at given momentum are blocked. It is asymmetric in $\bm{p}$ for $p_{\textrm{min}}<p<p_{\textrm{max}}$ (above $p_{\textrm{max}}$ all transitions are allowed, below $p_{\textrm{min}}$ all are blocked). Frequencies $\omega_{\textrm{min}}<\omega<\omega_{\textrm{max}}$ correspond to transitions for which only part of the spectrum is Pauli blocked.}
\label{minmaxfreq1}
\end{figure}

Below we focus on the free carrier contribution, $\sigma^{\textrm{(free)}}_{\alpha\beta}(\omega)$ in Eq.~\eqref{eq:sigma_free}.
Since  the  occupation factor in this term, $\left[n_+(\bm{p})-n_-(\bm{p})+1\right]$, is nonzero only close to the Weyl nodes,   $\sigma^{\textrm{(free)}}_{\alpha\beta}(\omega)$ may be written as a sum of partial contributions of individual Weyl nodes, $\sigma^{(n)}_{\alpha \beta}(\omega)$, see Eqs.~\eqref{eq:sigma_free_l} and~\eqref{eq:sigmaofp} below.
The latter may be expressed in terms of the Fermi energy in the node, $\mu$, and the parameters of the  Weyl Hamiltonian describing the electron dynamics near the node,
\begin{align}
\label{Hamilton}
\mathcal{H}(\bm{p})=\bm{u}\cdot\bm{p}\ \sigma_0+\ v_{ij} p_i\sigma_j.
\end{align}
Here  $\sigma_{\mu}$ are the Pauli matrices, the velocity matrix $v_{ij}$ is real symmetric, and the velocity $\bm{u}$ describes the  tilt of the energy cone
~\cite{footnote_neutrality}.
The index $n$ labelling the nodes has been omitted in these expressions.
The corresponding energy spectrum, $E_{\pm}(\mathbf{p})=\bm{u}\cdot\bm{p}\pm \sqrt{\sum_j(v_{ij}p_i)^2}$ is shown in Fig. \ref{minmaxfreq1}.

Before proceeding with quantitative consideration we discuss qualitatively the physical origin of the nonvanishing anomalous Hall response $\sigma^{\textrm{(free)}}_{xy}(\omega)$.
First we note that a finite Hall conductivity of an individual Weyl node is allowed by symmetry. It corresponds to the antisymmetric part of the conductivity tensor, $\sigma_{\alpha \beta}$, and is proportional to $\varepsilon_{\alpha \beta \gamma} u_\gamma$.
Here the tilt velocity $\bm{u}$ breaks the time reversal symmetry of a given node, and the Levi-Civita tensor $\varepsilon_{\alpha \beta \gamma}$ is provided by the chirality of the Weyl node (which is given by $\chi=\mathrm{sgn} (\det v)$).
 Although the above argument applies to both time-reversal invariant and noninvariant WSMs in the former the Hall contributions of Weyl nodes related by TR symmetry cancel each other. Indeed, the Hamiltonian of a TR partner may be obtained from Eq.~\eqref{Hamilton} by making the substitution $\bm{u} \to -\bm{u}$. This does not change the chirality of the node but changes the sign of the Hall response.
 In TR breaking WSMs such a cancellation does not occur and the Hall responses of individual nodes do not sum up to zero. Similarly, the responses of nodes linked by inversion symmetry generally add up as both $\bm{u}$ and $\chi$ change sign.

It is useful to see how a nonvanishing free carrier Hall response arises in the framework of Eq.~\eqref{eq:sigma_free}.
Note that tilting the energy dispersion by $\bm{u}$ amounts to a mere energy shift of the two-state system defined at each $\bm{p}$ and thus affects neither the matrix elements nor the energy denominators, i.e. $\sigma_{\alpha\beta}(\omega,\bm{p})$ is independent of the tilt.
In the absence of $\bm{u}$ the only  vector breaking time reversal symmetry is the momentum $\bm{p}$. Thus by the Onsager symmetry principle, $\sigma _{\alpha \beta}(\omega, \bm{p})= \sigma _{ \beta \alpha}(\omega, -\bm{p})$, the Hall response  must be odd in $\bm{p}$. It might seem that upon the integration over momentum this would give a vanishing result~\citep{burkovAHE14}, however this is not the case.
The occupation factor in Eq.~\eqref{eq:sigma_free} depends on the tilt velocity $\bm{u}$, which makes it asymmetric in $\bm{p}$, see Fig.~\ref{minmaxfreq1}. As a result the momentum sum in Eq.~\eqref{eq:sigma_free} is nonzero.
Physically the nonvanishing Hall response of free carriers arises due to asymmetric in $\bm{p}$ Pauli blocking of the filled band response. In a similar fashion asymmetric blocking leads to photocurrents in WSMs \cite{photoWSM17}.

Note that the occupation factor is asymmetric in $\bm{p}$ if and only if the node is tilted and the Fermi energy does not lie exactly at the nodal point. It is asymmetric in the region of momenta $p\in[p_{\textrm{min}},p_{\textrm{max}}]$ with $p_{\textrm{min/max}}=\lvert\mu\rvert/v_f(1\pm U)$ (here U is the magnitude of the tilt velocity in units of the Fermi velocity and for simplicity $v_{ij}=\chi v_f\delta_{ij}$). 

Let us now proceed with quantitative consideration.
For brevity we set $\hbar=c=1$ in intermediate steps.
The Kubo formula relates the optical  conductivity to the retarded current-current correlation function
\begin{equation}
\sigma_{\alpha\beta}(\omega)=\frac{i}{\omega}\Pi^{\textrm{Ret}}_{\alpha \beta}(\omega).
\end{equation}
The retarded correlator $\Pi^{\textrm{Ret}}_{\alpha \beta}(\omega)$ is obtained from the
Matsubara  current-current correlation function
\begin{equation}
\label{eq:Pi_Matsubara}
\Pi_{\alpha \beta}(i\omega_n)=\frac{T}{V} \sum_{m,\bm{p}} \textrm{tr}\big[j_{\alpha}\mathcal{G}(i\epsilon_m+i\omega_n,\bm{p})j_{\beta}\mathcal{G}(i\epsilon_m,\bm{p})\big]
\end{equation}
by analytic continuation to real frequencies,  $i\omega_n\rightarrow\omega+i0=\omega_+$. In Eq.~\eqref{eq:Pi_Matsubara} $T$ denotes temperature, $V$ volume
and the trace is taken over the spinor indices. The Matsubara Green function corresponding to Hamiltonian \eqref{Hamilton} is
\begin{align}
\mathcal{G}(i\epsilon_n,\bm{p})&=\frac{(i\epsilon_n-\bm{u}\cdot\bm{p})\sigma_0+\ v_{ij} p_i\sigma_j}{\left( i\epsilon_n-E_{+}(\bm{p})\right)\left( i\epsilon_n-E_{-}(\bm{p})\right)},
\end{align}
and the current operator is given by
\begin{align}
j_{\alpha}=-\frac{\delta}{\delta A_{\alpha}}\mathcal{H}(\bm{p}-e\bm{A})=e(u_{\alpha}\sigma_0+v_{\alpha j}\sigma_j).
\end{align}
To simplify notation we rescale momenta, $l_j=v_{ij} p_i$, and the tilt velocity, $\chi U_i=v_{ij}^{-1}u_j$, where we have factored out the chirality in order to consider the effect of tilt and chirality separately. Performing the frequency summation in Eq.~\eqref{eq:Pi_Matsubara} and subtracting the filled band contribution as explained above, we obtain the free carrier contribution to the conductivity from an individual node,
\begin{equation}\label{eq:sigma_free_l}
  \sigma^{\textrm{(n)}}_{\alpha\beta}(\omega)= \sum_{\bm{l}}\sigma^{(n)}_{\alpha\beta}(\omega,\bm{l})\left[n^{(n)}_{+}(\bm{l})-n^{(n)}_{-}(\bm{l})+1\right].
\end{equation}
In this expression  $n^{(n)}_{\pm}(\bm{l})=n_f(\chi\bm{l}\cdot\bm{U}\pm l-\mu)$ and
\begin{equation}\label{eq:sigmaofp}
\sigma^{(n)}_{\alpha\beta}(\omega,\bm{l})=\frac{1}{V}\frac{2ie^2v_{\alpha i} v_{\beta j}}{\lvert\det v \rvert}\,\frac{2l^2\delta_{ij}-2l_il_j+i\omega_+\varepsilon_{ijk}l_k}{\omega l(4l^2-\omega_+^2)},
\end{equation}
where the sum now is over rescaled momenta $\bm{l}$.
The symmetry arguments discussed above are manifest in this expression: the Hall (antisymmetric in $\alpha \beta$) response arises from the term  $\propto \varepsilon_{ijk}$, and exists only  if  $\bm{U}$ is nonzero,  since otherwise the occupation factor is symmetric under $l_{k}\rightarrow-l_k$.
Changing $\bm{l}\rightarrow\chi\bm{l}$ we observe that only the antisymmetric term is sensitive to the chirality of the node. The same applies to change of sign of the tilt velocity. Moreover,  we see that Hall response arises even for the isotropic velocity matrix $v_{ij} \propto \delta_{ij}$; the anisotropy of $v_{ij}$ merely amounts to anisotropic rescaling. Therefore  the contribution of valley $n$ to the optical conductivity tensor may be expressed in the form
\begin{equation}
\label{eq:sigma_tilde}
\sigma^{(n)}_{\alpha\beta}(\omega)=\frac{v_{\alpha i} }{v_f} \tilde{\sigma}^{(n)}_{ij}(\omega) \frac{v_{j \beta} }{v_f},
\end{equation}
where $v_f^3=\lvert\det v\rvert$ and $\tilde{\sigma}^{(n)}_{ij}(\omega)$ is the response corresponding  to the isotropic case,
 $v_{ij}=\chi v_f\delta_{ij}$. It is useful to note that after rescaling $\bm{U}$ is the only available vector breaking rotational invariance in the single-node problem. This allows to express the components of $\tilde{\sigma}$ in terms of universal functions $f^{(n)}$ depending only on the parameters of node $n$ such that
 \begin{subequations}\label{universal_sigma}
   \begin{eqnarray}
\tilde{\sigma}^{(n,\textrm{Hall})}_{ij}(\omega)&=&\varepsilon_{ijk}\hat{U}_k f^{(n)}_{\textrm{Hall}}(\omega),\\
\tilde{\sigma}^{(n,\perp)}_{ij}(\omega)&=& \left(\delta_{ij}-\hat{U}_i\hat{U}_j\right)f^{(n)}_{+}(\omega),\\
\tilde{\sigma}^{(n,\parallel)}_{ij}(\omega)&=&\hat{U}_i\hat{U}_j f^{(n)}_{-}(\omega),
\end{eqnarray}
 \end{subequations}
where $\hat{U}=\bm{U}/U,\ U=\lvert \bm{U}\rvert$. We assume $U<1$, i.e. only consider type-I WSMs~\cite{footnote_typeII}. As we wish to obtain closed form solutions we take the zero temperature limit. For details of the calculation see the supplemental material.  Restoring $\hbar$ we obtain for the frequency dependence of the free carrier conductivity of an individual node $n$
\begin{align}
\label{eq:free_hall}
f^{(n)}_{\textrm{Hall}}(\omega) = &\frac{-\chi \textrm{sgn}(\mu)e^2}{16\pi^2\hbar v_f U^2 } \times \nonumber  \\
&\left[\lvert\mu\rvert (L_1 + 2U)+\left(\frac{1-U^2}{4}\omega+\frac{\lvert\mu\rvert^2}{\omega}\right)L_2 \right]
\end{align}
and
\begin{eqnarray}\label{eq:xxzz}
f^{(n)}_{\pm}(\omega)&=&\frac{ie^2}{16\pi^2\hbar v_fU^3}\Bigg\{\frac{\lvert\mu\rvert^2a_{\pm}}{\omega}\left[\frac{4U(2+ a_{\pm}U^2)}{3(1-U^2)}+L_1\right]\nonumber \\
&&+\omega\left[\frac{U^3}{3}L_3+ a_{\pm}\left(\frac{1}{12}\pm\frac{U^2}{4}\right)L_1\right] \nonumber \\
&&+ a_{\pm}\left[\left(1\pm U^2\right)\frac{\lvert\mu\rvert}{2}+\frac{2\lvert\mu\rvert^3}{3\omega\omega_+}\right]L_2\Bigg\}.
\end{eqnarray}
In Eqs.~\eqref{eq:free_hall} and \eqref{eq:xxzz} we introduced the notation $a_{\pm}=-1/2\pm3/2$ and
\begin{subequations}\label{eq:L12}
  \begin{eqnarray}
L_1 &=& \ln
\frac{\omega^2_{\textrm{min}}-\omega^2_+}{\omega^2_{\textrm{max}}-\omega^2_+}, \\
L_2 &=& \ln\frac{(\omega_++\omega_{\textrm{min}})(\omega_+-\omega_{\textrm{max}})}{(\omega_+-\omega_{\textrm{min}})(\omega_++\omega_{\textrm{max}})}, \\
L_3 &=&\ln\frac{\omega^2_{\textrm{min}}-\omega^2_+}{-\omega^2_+}
+\ln\frac{\omega^2_{\textrm{max}}-\omega^2_+}{-\omega^2_+},
\end{eqnarray}
\end{subequations}
and also  $\omega_{\textrm{min/max}}=2\lvert\mu\rvert/(1\pm U)$ for the limiting frequencies of partially blocked transitions, c.f. Fig.~\ref{minmaxfreq1}.

\begin{figure}[t!]
\centering
    \includegraphics[width=0.99\columnwidth]{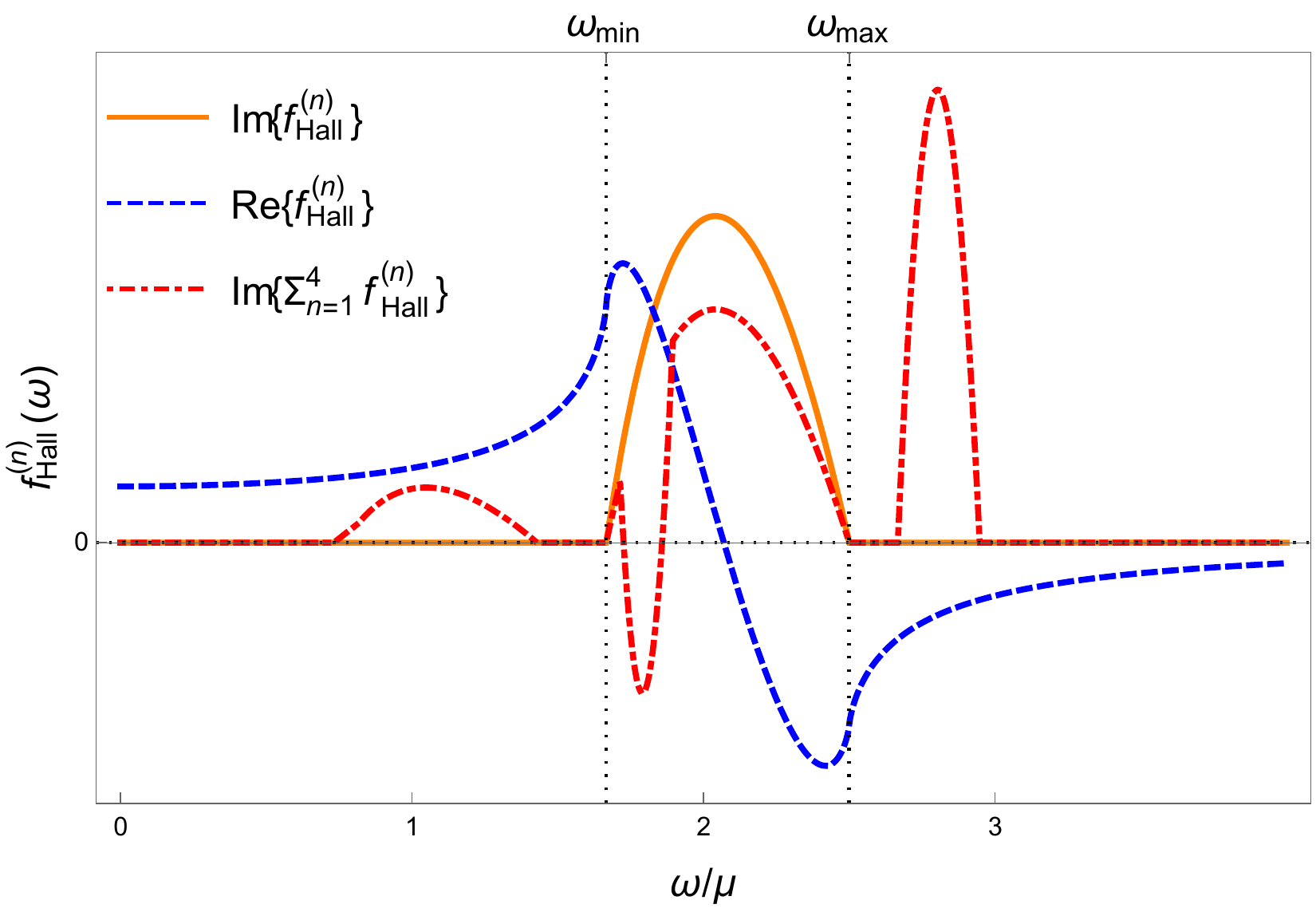}
\caption{\footnotesize Frequency dependence of the free carrier AH conductivity for a single node with $U=0.2$ and $\mu=v_f=1$. The imaginary part (solid orange) is non-zero only in the intervals $\omega_{\textrm{min}}<\omega<\omega_{\textrm{max}}$. The width of this region is determined by tilt and chemical potential. The dot-dashed red graph shows the response of a system of four nodes with  $\mu=(0.5,0.9,1,-1.4)$ and tilts $U=(0.3,0.05,0.2,0.05)$. The nodes with $\mu=0.9, \ 1.4$ have $\chi=-1$. Here, charge neutrality was ignored for simplicity.}
\label{condHall1}
\end{figure}

Equations \eqref{eq:sigma_tilde} - \eqref{eq:L12} are the central results of this paper~\cite{footnote_pockets}. In particular Eq.~\eqref{eq:free_hall} gives the frequency dependence of the free carrier contribution to the AHE, which is depicted in Fig.~\ref{condHall1}.
The imaginary part arises from real  optical transitions that are asymmetrically Pauli-blocked. It exists only in the frequency interval $\omega_{\textrm{min}}<\omega<\omega_{\textrm{max}}$.
The real part exhibits a resonant structure at frequencies $\pm\omega_{\textrm{min/max}}$. The red and dot-dashed graph shows the imaginary part of the free carrier Hall conductivity in a four node system given by the sum of individual node contributions.

\begin{figure}[t!]
\centering
    \includegraphics[width=0.99\columnwidth]{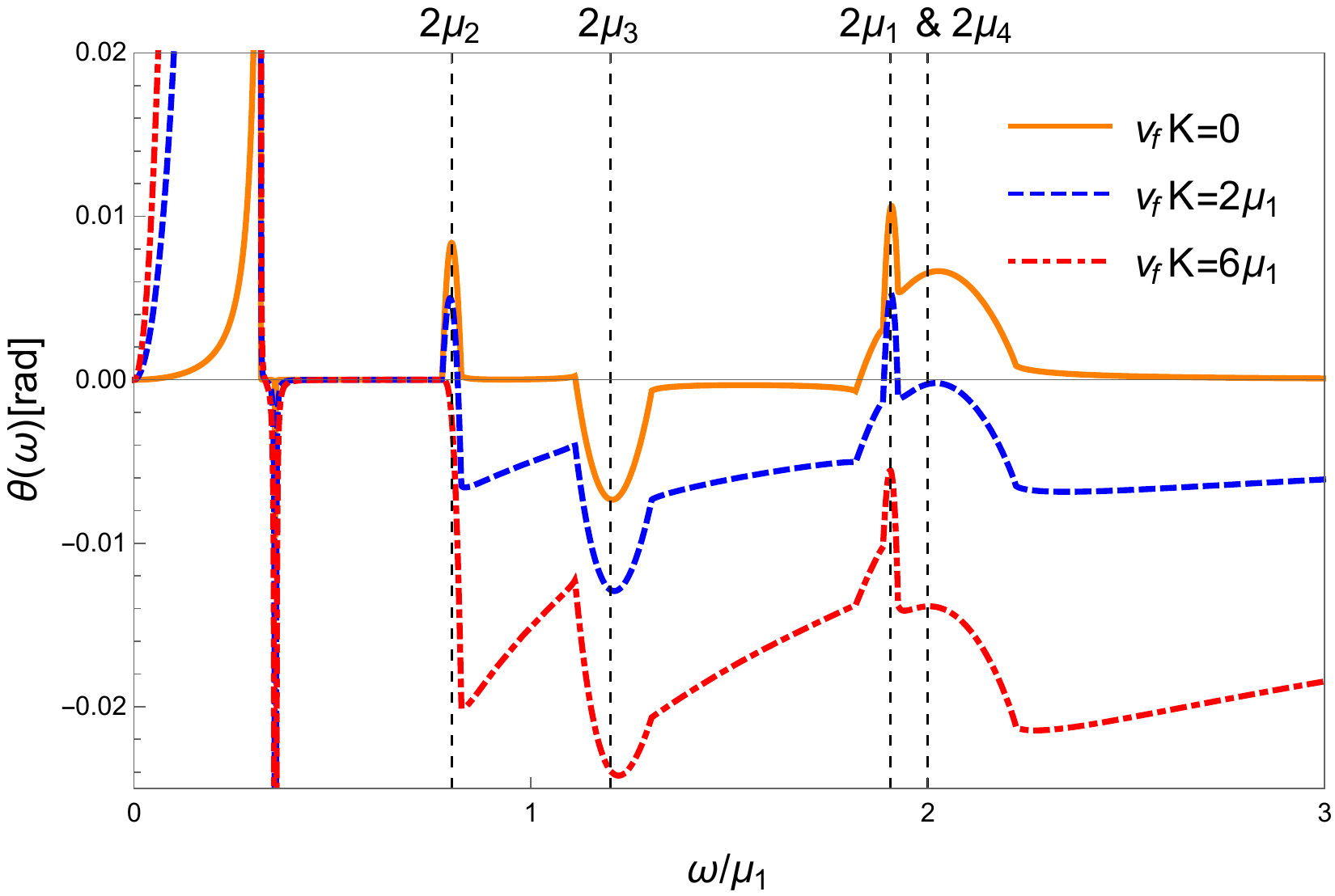}
\caption{\footnotesize Kerr spectra for a generic TR broken system of four nodes not related by symmetry. Charge neutrality is enforced. The AH response is modelled using the free carrier contribution plus the dc universal contribution $\sigma_{xy}^{\textrm{(u)}}=e^2K/2\pi h$. We used parameters $\epsilon_{\infty}=5$ and $\alpha'=e^2/\hbar v_f=1$. Each node contributes a peak to the spectrum. The features at low frequency are due to the plasmon mode. As long as $\omega_p<\omega_{\textrm{min}}$ for the lowest lying free carrier resonance and the universal component is small, the free carrier features are mostly determined by the imaginary part of the Hall conductivity. Large $\sigma_{xy}^{\textrm{(u)}}$ suppresses the peaks but kinks remain at $\omega=\omega_{\textrm{min/max}}$.}
\label{kerrangleManyCones}
\end{figure}

A nonzero ac Hall conductivity gives rise to the Kerr effect. It can be shown that the contribution of surface arc states is small in $v/c$~\cite{footnote_arcstates}. As a result the Kerr effect is described by the bulk optical conductivity. In particular,  in the polar Kerr effect geometry the angle of rotation of the polarization plane of reflected light is given by \cite{bookKerr}
\begin{equation}\label{eq:kerr}
\theta(\omega)=\textrm{Im}\left[\frac{4\pi\sigma_{xy}(\omega)}{\omega\sqrt{\epsilon_{xx}(\omega)}(\epsilon_{xx}(\omega)-1)}\right].
\end{equation}

Below we discuss the implications of our results for the Kerr effect. For simplicity we assume the nodes to be isotropic, $v^{(n)}_{ij}=\chi^{(n)}v_n\delta_{ij}$, and all tilts to lie along the $z$-axis. Then, the full Hall conductivity that enters the Kerr response is $\sigma_{xy}(\omega)=\sigma^{\textrm{(u)}}_{xy}+\sum_n \sigma^{(n)}_{xy}(\omega)$ with $ \sigma^{(n)}_{xy}(\omega)$ given by Eqs.~\eqref{eq:sigma_tilde} to~\eqref{eq:free_hall}. Note that the frequency dependence of the band contribution has been neglected and only the dc universal component $\sigma_{xy}^{\textrm{(u)}}=e^2K/2\pi h$ was kept, where $K$ is the effective distance of nodes in momentum space (see Refs.~\cite{Haldane2004,Ran2011} for details).   This step is justified in the supplemental material.
The permittivity entering~\eqref{eq:kerr} is
\begin{equation}\label{eq:permittivity}
\epsilon_{xx}(\omega)=\epsilon_{xx}^{\textrm{(band)}}(\omega)+\frac{4\pi i}{\omega}\sum_{\textrm{nodes }n}\sigma^{(n)}_{xx}(\omega)
\end{equation}
with longitudinal free carrier conductivity of an individual node given by Eqs.~\eqref{eq:sigma_tilde} to~\eqref{eq:xxzz}.

Due to the gapless character of the electron spectrum the filled band contribution to the permittivity $\epsilon_{xx}^{\textrm{(band)}}(\omega)$  has a logarithmic frequency dependence at low frequencies of interest. Since this dependence arises from the low energy electron excitations described by the Weyl Hamiltonian \eqref{Hamilton} it may be described in terms of the  parameters of the Weyl nodes. A consideration similar to that of the free carrier contribution yields for isotropic valleys with Fermi velocities $v_n$
\begin{equation}\label{eq:perm_band}
\epsilon_{xx}^{\textrm{(band)}}(\omega)=\epsilon_{\infty}+\frac{e^2}{6\pi \hbar}\sum_{\textrm{nodes }n}\frac{1}{v_n}\ln\frac{\Lambda^2}{-\omega_+^2}.
\end{equation}
The frequency cut-off $\Lambda$ can be absorbed in the permittivity of inert bands $\epsilon_{\infty}$. For details see the supplemental material.

Figure \ref{kerrangleManyCones} shows a characteristic Kerr spectrum for a WSM with four nodes. The presence of free carriers at node $n$ leads to resonances in the Kerr angle which are (skewly) centered around $\omega=2\mu_n$ with width $U_n\mu_n/(1-U_n^2)$. The sign of the peaks is given by the product of chirality, projection of tilt along $\hat{e}_z$ and the sign of the chemical potential. The peak at low frequencies occurs at the plasmon frequency $\omega_p$ that corresponds to vanishing permittivity. If the plasmon response occurs at larger frequencies than free carrier features both free carrier and plasmon peak will remain present but change in shape.
As is clear from the graphs, the universal contribution to the AHE merely modifies the shape of the resonant features in the frequency dependence of the Kerr angle, while their locations are determined purely by the free carrier contribution. This allows to experimentally determine the doping level of individual valleys.

Note that extrapolation of our result for $f_{\textrm{Hall}}^{\textrm{(n)}}(\omega)$ to the \textit{dc} limit $\omega \to 0$ gives a finite result: assuming isotropy and $\bm{u}\parallel \hat{e}_z$,
\[
\sigma^{\textrm{(n)}}_{xy}(0)=f_{\textrm{Hall}}^{\textrm{(n)}}(0)=\frac{-\chi e^2\mu}{8\pi^2\hbar v_f}\left(\frac{2}{U}+\frac{1}{U^2}\ln\left[\frac{1-U}{1+U}\right]\right).
\]
This result is purely formal, as in the presence of impurities our results only apply in the collisionless regime $\omega \tau \gg 1$, where $\tau$ is the transport mean free time. Nevertheless, in the \textit{dc} regime $\omega \tau \ll 1$ the free carrier contribution to the AHE should remain finite. For high mobility conductors it is expected to be dominated by skew scattering of Weyl fermions and may be estimated as
\[
\sigma^{\textrm{(n,sk)}}_{xy}\propto\chi\frac{\tau^2}{\tau_{\textrm{sk}}}\frac{e^2\mu^2\eta(U)}{\hbar^2 v_f}.
\]
Here $1/\tau_{sk}$ is the skew scattering rate.
Note that  skew scattering is allowed by symmetry. For example, chirality allows to write the intranode skew scattering cross-section
 in the form $w_{kk'}\propto\bm{u}\cdot\left(\bm{k}\times\bm{k}'\right)$. It arises only beyond the lowest Born approximation for the scattering amplitude.
In this respect it is worth noting that in Ref. \citep{burkovAHE14} the effects of energy cone tilt on Pauli blocking and impurity skew scattering were not considered. This resulted in a vanishing free carrier contribution to the anomalous Hall conductivity.

In conclusion, we note that the developed microscopic theory of ac anomalous Hall conductivity, and its implications for the magneto-optical Kerr effect, apply not only to WSMs with spontaneously broken TR symmetry, but are also  relevant for systems in which the WSM phase is ``created"~\cite{trbreakingB16} by application of a magnetic field to TR-invariant Dirac semimetals. We would also like note that the diagonal components of the conductivity tensor have features at $\omega\sim 2\mu$. Thus, in TR-invariant Weyl semimetals the doping levels of individual valleys may be characterized by measuring the frequency dependence of the surface impedance.

\acknowledgments{
JFS gratefully acknowledges financial support by the PROMOS program of the DAAD (Deutscher Akademischer Auslandsdienst).   The work of AVA was supported by the U.S. Department of Energy   Office of Science, Basic Energy Sciences under Award No. DE-FG02-07ER46452. The work of DAP was supported by the National Science Foundation Grant No. DMR-1409089.}

\bibliography{biblioKERRupd,addrefs}

\begin{thebibliography}{41}
\bibitem[{foo({\natexlab{a}})}]{footnote_neutrality}
\bibinfo{note}{These parameters are not fully independent but are subject to
  the constraints set by the Nielsen-Ninomiya theorem, crystal symmetries and
  charge neutrality. Note that node $n$ contributes carrier density
  \begin{equation*}
  n(\mu_n,U_n,v_n)=\frac{2\mu_n^3}{4\pi^2\hbar^3v_n^3}\frac{1}{3(1-U_n^2)^2}.
  \end{equation*} For an undoped crystal the charge neutrality condition
  corresponds to $\sum_n n(\mu_n,U_n,v_n)=0$. Here
  $v_{\alpha\beta}=v_n\delta_{\alpha\beta}$ and $U$ is the absolute value of
  the tilt velocity in units of $v_n$}.
\bibitem[{foo({\natexlab{b}})}]{footnote_typeII}
\bibinfo{note}{The free carrier contribution to the conductivity of type-II WSM
  requires a different treatment since here the Fermi surface is not well
  captured by a single node model~\cite{zyuzin16}.}
\bibitem[{foo({\natexlab{c}})}]{footnote_pockets}
\bibinfo{note}{If in addition to the Weyl nodes there are further carrier
  pockets of different nature, these equations give the partial contribution
  due to the Weyl nodes.}
\bibitem[{foo({\natexlab{d}})}]{footnote_arcstates}
\bibinfo{note}{The expression in Eq.~\eqref{eq:kerr} assumes local response
  (absence of spatial dispersion) and ignores the contribution of the Fermi
  arcs. The first assumption is valid because the electron displacement during
  the oscillation period $\sim v/\omega$ is smaller than the skin depth in a
  Weyl semimetal $\delta =c/\sqrt{2\pi \omega \sigma_{xx} (\omega) } \sim
  c/\omega$. A straightforward calculation based on the Fermi golden rule shows
  that the contribution of arc states to the energy absorption rate is smaller
  than that of bulk states in $v/c$. Since the ratio of the Hall response to
  the diagonal response is independent of $v/c$ the arc state contribution to
  the Kerr effect is also small.}
\end{thebibliography}
\bibliographystyle{apsrev}

\end{document}